\documentclass{PoS}

\usepackage[utf8]{inputenc}
\usepackage{amsmath}
\usepackage{braket}
\usepackage{xcolor}
\usepackage{tikz}

\newcommand{\GeV}{\,\text{GeV}}
\newcommand{\TeV}{\,\text{TeV}}
\newcommand{\TR}{T_\text{R}}

\newcommand{\beq}{\begin{equation}}
\newcommand{\eeq}{\end{equation}}
\newcommand{\lagr}{\mathcal{L}}
\newcommand{\hc}{\textnormal{h.c.}}

\newcommand{\mhu}{\ensuremath{m_{H_u}}}
\newcommand{\stau}{{\widetilde{\tau}}}

\newcommand{\Eqref}[1]{eq.~\eqref{#1}}

\newcommand{\Figref}[1]{fig.~\ref{#1}}

\newcommand{\LOSP}{NLSP}

\definecolor{darkgreen}{RGB}{0,170,0}
\definecolor{ourpurple}{RGB}{135,0,130}

\title{Gaugino Mediation with Large Trilinears}

\ShortTitle{Gaugino Mediation with Large Trilinears}

\author{Jan Heisig\\
        RWTH Aachen University\\
        E-mail: \email{heisig@physik.rwth-aachen.de}}

\author{\speaker{Jörn Kersten}\\
        University of Bergen\\
        E-mail: \email{joern.kersten@uib.no}}

\author{Nick Murphy\\
        CP${\:\!}^3$-Origins, University of Southern Denmark\\
        E-mail: \email{murphy@cp3.sdu.dk}}

\author{Inga Strümke\\
        University of Bergen\\
        E-mail: \email{inga.strumke@uib.no}}

\abstract{
Gaugino mediation is an attractive supersymmetry breaking scheme
naturally avoiding flavor problems by suppressing the soft sfermion
masses at a high-energy scale. We consider an extension of the original
model which yields non-vanishing trilinear scalar couplings. This
increases the viable parameter space predicting a sufficiently large
Higgs mass. Assuming the gravitino to be the lightest superparticle, we
consider additional constraints from direct searches at the LHC, finding
allowed points with a neutralino, sneutrino or stau next-to-lightest
superparticle.
}

\FullConference{Corfu Summer Institute 2017 ``School and Workshops on
Elementary Particle Physics and Gravity''\\
		 2--28 September 2017\\
		 Corfu, Greece}

\begin{document}

\section{Introduction}
While low-energy supersymmetry (SUSY) remains the most elegant solution
of the hierarchy problem, it is being pressured on a number of fronts.
In addition to the classical problems like the flavor and gravitino
problem, the lack of a signal for superparticles at the LHC becomes an
increasingly severe issue, which might be dubbed the SUSY discovery
problem.  This may imply that SUSY is not realized in nature after all.
However, it may also imply that the realization of SUSY chosen by nature
has somewhat unusual features that limit the effectiveness of the LHC
searches.

One such feature is the nature of the lightest SUSY particle (LSP).  If
it is the neutralino, as is usually assumed, the gravitino is unstable.
Due to its extremely weak interactions, it has a relatively long
lifetime of up to several years.  In this case, the energetic decay
products created by gravitino decays in the early universe destroy
nuclei produced by Big Bang Nucleosynthesis (BBN)
\cite{Falomkin:1984eu,Khlopov:1984pf,Ellis:1984eq}.  The observed
abundances of primordial light elements therefore either require a
gravitino mass $m_{3/2} \gg 1\TeV$ or a reheating temperature after
inflation $\TR \lesssim 10^6\GeV$ \cite{Kawasaki:2017bqm}.  The former
constraint implies a quite unnatural mass spectrum in most scenarios of
SUSY breaking, whereas the latter one prevents thermal leptogenesis
without fine-tuning \cite{Buchmuller:2002rq}.  This motivates scenarios
where the gravitino is the LSP and thus stable.  For SUSY at the TeV
scale and $\TR \sim 10^9\GeV$, thermal production shortly after
inflation yields a gravitino density that is consistent with the
observed dark matter density, if the gravitino has a mass of some tens
of GeV \cite{Bolz:2000fu,Pradler:2006qh}.  Thus, a relatively heavy
gravitino LSP is a viable cold dark matter candidate.  In this case the
next-to-LSP (NLSP) becomes long-lived in the absence of R-parity
violation, as the only superparticle it can decay into is the gravitino
with its superweak interactions.  So even in this scenario we have to
worry about the effect of late decays on BBN\@.
For NLSP masses below a TeV, this rules out neutralinos,
charginos and gluinos as
NLSPs, but a slepton or stop NLSP can be allowed.  Charged NLSPs are
further constrained since they form bound states with nuclei, which
alters BBN reaction rates \cite{Pospelov:2006sc}.  Consequently, the
gravitino problem is present also in gravitino LSP scenarios but
significantly alleviated.  In addition, the LHC phenomenology is
determined by the properties of the NLSP and quite different from the
standard neutralino LSP case, which may alleviate the discovery problem.

The appearance of unacceptably large flavor and CP violation in a
generic SUSY scenario is a problem that can be tackled by the mechanism
mediating SUSY breaking from the hidden to the visible sector.  Here we
will consider the example of gaugino mediation
\cite{Kaplan:1999ac,Chacko:1999mi}.  It employs a setup with extra
spacetime dimensions to prevent direct couplings between the sfermions
and the field breaking SUSY, which suppresses the soft sfermion masses
at a high-energy scale and thus avoids flavor problems.
It also allows a gravitino LSP with a mass in the range needed to
alleviate the gravitino problem \cite{Buchmuller:2005rt} and a slepton
NLSP \cite{Buchmuller:2005ma}.  Alternatively, the lightest neutralino
could be the LSP and dark matter particle, at the price of a heavy
gravitino to satisfy BBN constraints.

In the realization that was proposed originally, gaugino mediation
also yields suppressed trilinear scalar couplings, which is
unfortunate since the measured Higgs mass \cite{Aad:2015zhl}
then requires a unified gaugino mass of $m_{1/2} \gtrsim 3\,$TeV
and thus very heavy superparticles \cite{Kitano:2016dvv}.
However, a simple extension of the model produces non-vanishing
trilinears and thus allows for a lighter superparticle spectrum
\cite{Brummer:2012ns,Heisig:2017lik}.  
In the following, we will explore the phenomenology of this extended
scenario, summarizing the results of \cite{Heisig:2017lik}.


\section{Trilinear-Augmented Gaugino Mediation}\label{sec:breaking}
\subsection{General Setup}
\begin{figure}[bthp]
\centering
\vspace{-2mm}
\begin{tikzpicture}
\draw[color=blue,thick] (0,0)--(1.4,1.4)--(1.4,5)--(0,3.6)--(0,0);
\path (-0.5,1.5) -- node[sloped,xslant=.5,color=violet] {$S$} (2,3.5);
\draw[color=blue,thick] (3.5,0)--(4.9,1.4)--(4.9,5)--(3.5,3.6)--(3.5,0);
\path (3.2,1.5) -- node[sloped,xslant=.5] {\parbox{1.3cm}{\centering\color{blue}MSSM\\matter}} (5.2,3.5);
\node at (3.1,4.9) {\color{blue}$4D$ branes};
\node at (2.45,2.5) {\parbox{2cm}{\centering\OliveGreen $W$\\$H_u,H_d$\\\Orange $G$}};
\draw[->,thick] (-0.5,0)--(5.5,0);
\node at (5.2,-0.25) {\small Extra dimension};
\draw (0,0)--(0,-0.1);
\node at (0,-0.25) {\footnotesize $y_S$};
\end{tikzpicture}
\vspace{-2mm}
\caption{Geometrical setup of gaugino-mediated SUSY breaking ($S$:
SUSY-breaking field, $W$: gauge supermultiplets, $G$: gravity
supermultiplet).}
\label{fig:setup}
\end{figure}
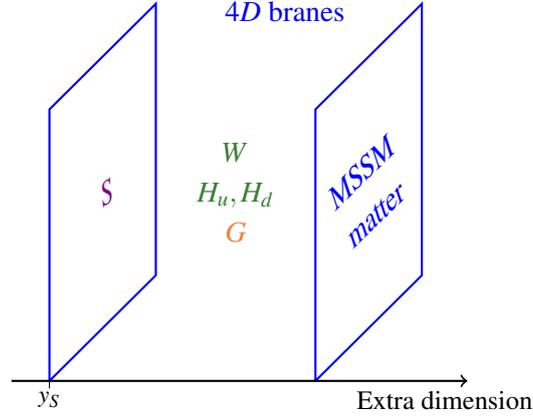

The setup of gaugino mediation \cite{Kaplan:1999ac,Chacko:1999mi} is illustrated in \Figref{fig:setup}.
There are $D$ spacetime dimensions, $D-4$ of which are compact with
volume $V_{D-4}$.  This volume determines the compactification scale
$M_c \equiv (1/V_{D-4})^{\frac{1}{D-4}}$, which we will assume to equal
the unification scale of about $10^{16}\GeV$.  The $D$-dimensional bulk
contains two $4$-dimensional branes.  The MSSM matter fields are
localized on one of them (MSSM brane), while the SUSY-breaking sector is
localized on a different brane (hidden brane).  For our purposes it is
sufficient to consider a single chiral superfield $S$ of the hidden
sector, which is a Standard Model (SM) singlet and develops an $F$-term
vacuum expectation value (VEV) $\braket{F_S}$ that breaks SUSY\@.

The gauge fields, the graviton and the gravitino are placed in the bulk.
Therefore, gauginos and the gravitino can couple to the SUSY-breaking
field localized on the hidden brane and obtain soft masses proportional
to $\braket{F_S}$.  In contrast, the SM fermions and their superpartners
are constrained to the MSSM brane.  Consequently, the sfermions'
couplings to $S$ and the corresponding soft masses are strongly
suppressed.  This avoids contributions to flavor violation from the
scalar soft mass matrices and thus solves a part of the SUSY flavor
problem.  The remaining part of the problem, potential flavor violation
from trilinear scalar couplings, will be addressed below.

In the version of gaugino mediation proposed in \cite{Kaplan:1999ac},
also the Higgs superfields are localized on the MSSM brane and thus
do not obtain soft masses, whereas the version of \cite{Chacko:1999mi}
features bulk Higgses and non-vanishing Higgs-$S$ couplings.
We will consider the latter setup in the following.

\subsection{Trilinear Couplings}

The most general $D$-dimensional Lagrangian coupling bulk Higgses to the
SUSY-breaking field is
\begin{multline} \label{eq:dlmssm}
\lagr_{HS} =
\frac{\delta^{(D-4)}(y-y_S)}{M^{D-4}} \left[
\frac{S}{M} \left(a\hat H_u^\dagger \hat H_d^\dagger+b_u \hat H_u^\dagger \hat H_u+b_d\hat H_d^\dagger \hat H_d\right)+\hc
+{}\right.
\\
\left.{}+
\frac{S^\dagger S}{M^2} \left(c_u\hat H_u^\dagger \hat H_u+c_d \hat H_d^\dagger \hat H_d+(d \hat H_u \hat H_d+\hc)\right)
\right]_D + \dots \,,
\end{multline}
where $y_S$ is the position of the hidden brane in the extra
dimensions and hats denote bulk fields with canonically normalized
kinetic terms in $D$ dimensions.  The dots refer to terms containing
additional powers of $S$, which do not lead to qualitatively new results
but at most to corrections suppressed by powers of $M$, the scale up to
which the theory is valid.
Finally, $a$, $b_{u,d}$, $c_{u,d}$ and $d$ are dimensionless couplings.

Importantly, the original gaugino mediation models
\cite{Kaplan:1999ac,Chacko:1999mi} did not include the terms with
couplings $b_u$ and $b_d$.  It is precisely these terms that give rise
to trilinear couplings \cite{Brummer:2012ns}.  In order to show this,
let us integrate over the extra dimensions to arrive at the effective
$4$-dimensional Lagrangian valid below $M_c$,
\begin{multline} \label{eq:4lmssm}
\lagr_{4D} \supset
\left[ H_u^\dagger H_u + H_d^\dagger H_d \right]_D +
\rho
\left[\frac{S}{M}\left(a H_u^\dagger H_d^\dagger+b_u H_u^\dagger H_u+b_d H_d^\dagger H_d\right)+\hc
+{}\right.
\\
\left.{}+
\frac{S^\dagger S}{M^2}\left(c_u H_u^\dagger H_u+c_d H_d^\dagger H_d+(d H_u H_d+\hc)\right)
\right]_D ,
\end{multline}
where we have denoted the canonically normalized zero modes of the Higgses
by $H_{u,d}$ and included their kinetic terms.  The canonical
normalization of the bulk fields has yielded a volume factor
\begin{equation}
\rho \equiv
\frac{1}{V_{D-4} M^{D-4}} \equiv
\left(\frac{M_c}{M}\right)^{D-4}
\end{equation}
in front of the bulk-brane interaction terms.
The terms proportional to $b_u$ and $b_d$ can be removed by the field
redefinitions%
\footnote{The factor $\rho$ in \Eqref{eq:Redef} is missing in
\cite{Heisig:2017lik}.}
\begin{equation} \label{eq:Redef}
H_{u,d} \equiv H'_{u,d} \left(1-\rho b_{u,d}\frac{S}{M}\right) ,
\end{equation}
which changes the Lagrangian \eqref{eq:4lmssm} to
\begin{multline} \label{eq:LTransformed}
\lagr_{4D} \supset
\left[ H_u'^\dagger H_u' + H_d'^\dagger H_d' \right]_D +
\rho
\left[\frac{S}{M}\left(a H_u'^\dagger H_d'^\dagger\right)+\hc
+{}\right.
\\
\left.{}+
\frac{S^\dagger S}{M^2}\left(c'_u H_u'^\dagger H_u'+c'_d H_d'^\dagger H_d'+
(d' H_u' H_d'+\hc)\right)
\right]_D
\end{multline}
(again omitting irrelevant higher-order terms), where
\begin{equation}
c'_{u,d} = c_{u,d} - \rho |b_{u,d}|^2
\quad,\quad
d' = d - \rho a^* \left( b_u+b_d \right) .
\end{equation}
Thus, we obtain a contribution to the $\mu$-term, soft Higgs masses and
$B\mu$ proportional to $\braket{F_S}$.
Gaugino masses are generated by a term in the gauge kinetic function
that we do not show here \cite{Kaplan:1999ac,Chacko:1999mi}.
In the part of the Lagrangian stemming from the superpotential 
$W_\text{MSSM} =
\bar{u} y_u Q H_u - \bar d y_d Q H_d - \bar e y_e L H_d + \mu H_u H_d$,
the field redefinitions \eqref{eq:Redef} yield
\begin{align}
\lagr_{4D} &\supset
\left[ \bar{u} y_u Q H'_u - \bar d y_d Q H'_d - \bar e y_e L H'_d
- \rho b_u \frac{S}{M} \, \bar{u} y_u Q H_u'
+ \rho b_d \frac{S}{M} \, \bar{d} y_d Q H_d'
+ \rho b_d \frac{S}{M} \, \bar{e} y_e L H_d'
\right]_F
\nonumber\\
&\supset
-\rho b_u \frac{\braket{F_S}}{M} \tilde u_\text{R}^* y_u \tilde Q_\text{L} H_u'
+\rho b_d \frac{\braket{F_S}}{M} \tilde d_\text{R}^* y_d \tilde Q_\text{L} H_d'
+\rho b_d \frac{\braket{F_S}}{M} \tilde e_\text{R}^* y_e \tilde L_\text{L} H_d'
\end{align}
and thus trilinear scalar couplings
\begin{equation} \label{eq:trilinears}
a_u = A_{u0} \, y_u \quad,\quad
a_d = A_{d0} \, y_d \quad,\quad
a_e = A_{d0} \, y_e
\end{equation}
with
\beq \label{eq:Afromb}
A_{u0} = \left(\frac{M_c}{M}\right)^{D-4} \frac{\braket{F_S}}{M} \, b_u
\quad,\quad
A_{d0} = \left(\frac{M_c}{M}\right)^{D-4} \frac{\braket{F_S}}{M} \, b_d \,.
\eeq
This result can also be derived from the general expressions for soft
SUSY-breaking terms in the supergravity formalism, see
\emph{e.g.}~\cite{Brignole:1997dp}, and by integrating out the Higgs
auxiliary fields \cite{Heisig:2017lik}.
As the trilinear and Yukawa matrices are proportional to each other,
they are simultaneously diagonalized when changing to the super-CKM
basis.  Consequently, the trilinears do not cause additional flavor
violation compared to the SM and the second part of the SUSY flavor
problem is solved, too.
Interestingly, the proportionality factors $A_{u0}$ for the up-type
squarks and $A_{d0}$ for the down-type squarks and charged sleptons can
be different.

\section{Phenomenology}\label{sec:pheno}
Let us now discuss the superparticle spectrum and the lightest Higgs
mass, as computed by \textsc{SPheno}~3.3.8~\cite{spheno,Porod:2011nf}
and
\textsc{FeynHiggs}~2.12.2~\cite{Heinemeyer:1998np,Heinemeyer:1998yj,Degrassi:2002fi,Frank:2006yh,Hahn:2013ria,Bahl:2016brp},
respectively, that can be obtained in
trilinear-augmented gaugino mediation.  From the above considerations it
follows that the free parameters are the gaugino masses, the Higgs soft
masses $m^2_{H_u}$ and
$m^2_{H_d}$, the trilinear couplings $A_{u0}$ and $A_{d0}$, as well as $B\mu$ and $\tan\beta$ at the
compactification scale.  The soft sfermion masses are negligibly small
at $M_c$.  Here we assume a unified gauge theory above the
compactification scale, resulting in a unified gaugino mass
$m_{1/2}$.  In addition, we restrict ourselves to the simplest case
regarding the trilinears, $A_{u0} = A_{d0} \equiv A_0$.  In this way we
arrive at a restricted realization of the NUHM2 scenario
\cite{Ellis:2002wv} with $m_0=0$.  Finally, we choose $\mu>0$,
$A_0\leq0$ and $m^2_{H_{u,d}}\geq0$.

\subsection{Higgs Mass}\label{sec:mh}

The observed value of the Higgs mass, $m_h = 125.09\GeV$ \cite{Aad:2015zhl},
is a challenge for SUSY model building, since it requires significant
corrections to the tree-level prediction $m_h < m_Z$.  The dominant
one-loop correction,
\begin{equation}
	\Delta m_h^2 \propto
	m_t^2 \left[ \log\frac{M_S^2}{m_t^2} +
	\frac{X_t^2}{M_S^2}
	\left( 1 - \frac{X_t^2}{12 \, M_S^2} \right)
	\right] ,
\end{equation}
depends on the overall superparticle mass scale
$M_S = \sqrt{m_{\tilde t_1} m_{\tilde t_2}}$
and the stop trilinear via
$X_t = A_t - \mu\cot\beta$.
As a small value of $M_S$ facilitates a discovery of SUSY, a large
(absolute) value of the trilinear coupling is desirable.  Consequently,
generating large trilinears in gaugino mediation is crucial for the
observability of the scenario.

As the theoretical uncertainty of Higgs mass calculations in the MSSM is
around $2\GeV$ \cite{Hahn:2013ria,Borowka:2014wla}, we consider
parameter space points with
$123\,\text{GeV} \leq m_h \leq 127\,\text{GeV}$ to be
allowed in our numerical analysis.
For each point we studied, the Higgs mass computed by
\textsc{FeynHiggs}~2.12.2 is typically about $3\GeV$ smaller than the
one found by \textsc{SPheno}~3.3.8.  However, a good agreement between
the results of \textsc{FeynHiggs} and \textsc{SPheno}~4 has been
reported \cite{Staub:2017jnp}.  Besides, we observed a downward shift of
the computed Higgs mass by about $1\GeV$ when switching from
\textsc{FeynHiggs}~2.11 to version 2.12, which was mainly caused by a
more accurate calculation of electroweak corrections to the
$\overline{\text{MS}}$ top mass \cite{Bahl:2016brp}.  As a result, it
seems likely that the superpartner mass scale required to reach the
observed Higgs mass had been underestimated in studies using older
versions of the codes.

\subsection{NLSP Candidates}\label{sec:spect}

\begin{figure}
\centering
\setlength{\unitlength}{1\textwidth}
\begin{picture}(0.98,0.42)
 \put(0.0,0.0){
  \put(0.07,0.03){\includegraphics[width=0.36\textwidth]{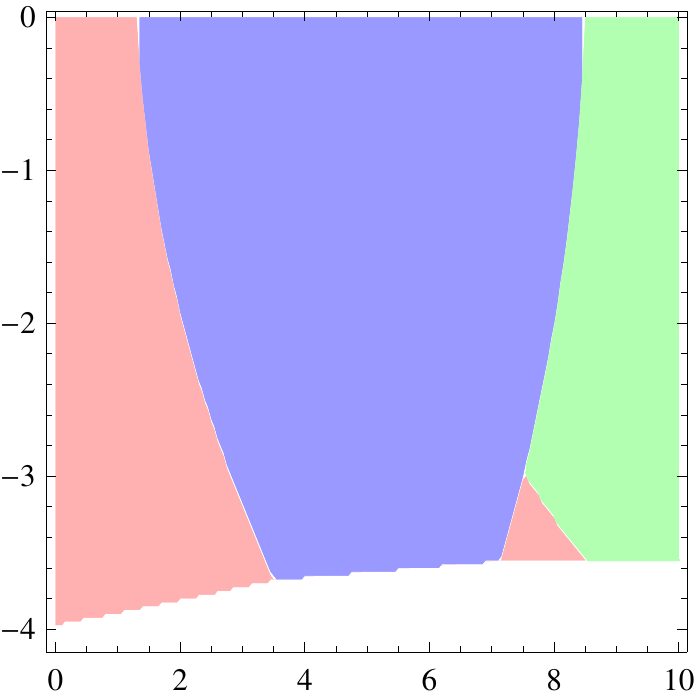}}
  \put(0.22,0.0){\footnotesize $m_{H_d}^2/m_{1/2}^2$}
  \put(0.03,0.18){\rotatebox{90}{\footnotesize $A_0/m_{1/2}$}}
  \put(0.24,0.25){\rotatebox{0}{\scriptsize  {\color{blue} $\chi^0$} }}
  \put(0.125,0.16){\rotatebox{0}{\scriptsize  {\color{red} $\widetilde \tau_\text{(R)}$} }}
  \put(0.335,0.11){\rotatebox{0}{\scriptsize  {\color{red} $\widetilde \tau_\text{(L)}$} }}
  \put(0.383,0.175){\rotatebox{0}{\scriptsize\OliveGreen $\widetilde\nu$}}
  \put(0.095,0.4){\rotatebox{0}{\footnotesize\color{black} $\tan\beta=10\;,\; m_{H_u}^2=0$}}
  \put(0.23,0.069){\rotatebox{0}{\tiny  Tachyonic spectrum}}
  }
 \put(0.48,0.0){
  \put(0.07,0.03){\includegraphics[width=0.36\textwidth]{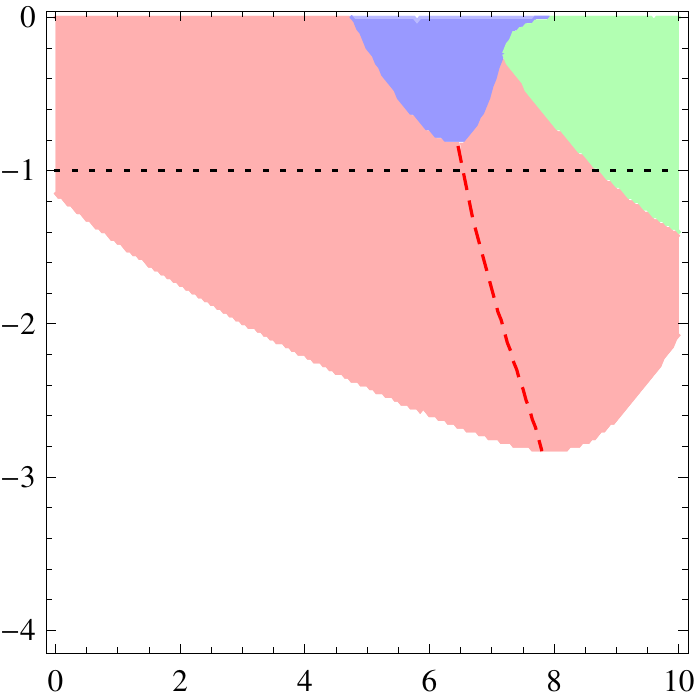}}
  \put(0.22,0.0){\footnotesize $m_{H_d}^2/m_{1/2}^2$}
  \put(0.03,0.18){\rotatebox{90}{\footnotesize $A_0/m_{1/2}$}}
  \put(0.293,0.354){\rotatebox{0}{\scriptsize  {\color{blue} $\chi^0$} }}
  \put(0.22,0.28){\rotatebox{0}{\scriptsize  {\color{red} $\widetilde\tau_\text{(R)}$} }}
  \put(0.36,0.25){\rotatebox{0}{\scriptsize  {\color{red} $\widetilde\tau_\text{(L)}$} }}
  \put(0.38,0.34){\rotatebox{0}{\scriptsize\OliveGreen $\widetilde\nu$ }}
  \put(0.095,0.4){\rotatebox{0}{\footnotesize\color{black} $\tan\beta=20\;,\; m_{H_u}^2=5\,\text{TeV}^2$}}
  \put(0.19,0.12){\rotatebox{0}{\tiny Tachyonic spectrum}}
  }
\end{picture}
\caption{Nature of the NLSP as a function of the down-type soft Higgs
mass and the trilinear scalar coupling for two different choices of
$\tan\beta$ and $m_{H_u}^2$.  In both panels, $m_{1/2}=2\TeV$, but
the figures look almost the same for different values of this parameter.
In the white regions at the bottom, a soft mass squared becomes
negative.  The red-dashed curve in the right panel indicates a maximally
mixed stau \LOSP, i.e., $\sin^2\theta_\stau = 1/2$.  Figures taken from
\cite{Heisig:2017lik}.
}
\label{fig:LSPplot}
\end{figure}
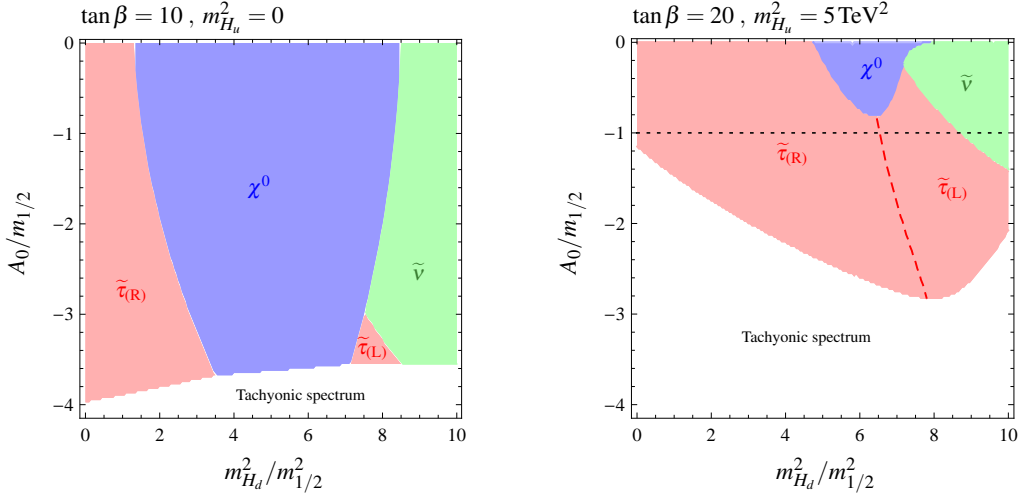

The superparticle mass spectrum is determined by the aforementioned input
parameters at the compactification scale and the renormalization group
running to low energy.  The overall SUSY \emph{mass scale} is determined
by the unified gaugino mass $m_{1/2}$.  To a good approximation, a
change of this parameter results in a common rescaling of all
superparticle masses.  The \emph{mass ordering} depends on the other
free parameters, among which the soft Higgs mass $m^2_{H_d}$ and the
trilinear coupling $A_0$ are the most important.  Assuming a gravitino
LSP, the lightest of the MSSM superparticles is the NLSP\@.  For
$m^2_{H_d}=0$, this particle is always the (predominantly right-handed)
lighter stau.  For non-trivial values of $m^2_{H_d}$ and $A_0$, the NLSP
can also be a neutralino or a tau sneutrino, as illustrated in
\Figref{fig:LSPplot}.  For non-zero $A_0$, it is even possible to obtain a
stau NLSP that is a maximal mixture between $\widetilde\tau_\text{R}$
and $\widetilde\tau_\text{R}$.  This happens, for example, for the
points on the long-dashed line in the right panel of the figure.

\subsection{LHC Constraints} \label{sec:collider}

As the gravitino cannot be lighter than about $10\GeV$ in gaugino
mediation \cite{Buchmuller:2005rt}, the NLSP is effectively stable on
timescales relevant for collider experiments.  In the stau NLSP
case, the scenario is therefore tested by searches for heavy stable
charged particles (HSCP) at the LHC\@.  Performing a Monte Carlo
simulation of the expected signal with
\textsc{MadGraph5\_aMC@NLO}~\cite{Alwall:2014hca} (event generation) and
\textsc{Pythia}~6~\cite{Sjostrand:2006za} (total cross section, decay,
showering, hadronization), we applied a CMS search with an integrated
luminosity of $18.8\,\text{fb}^{-1}$ at the $8\TeV$ run of the
LHC~\cite{Khachatryan:2015lla} to constrain the parameter space.

We also considered the latest available $13\TeV$ results, a preliminary
CMS analysis using $12.9\,\text{fb}^{-1}$~\cite{CMS-PAS-EXO-16-036}.  As
fewer details of the analysis are provided than for the run-$1$ search,
a reinterpretation of the results is more difficult.  Nevertheless, we
obtained a meaningful estimate of the exclusion bound, which is slightly
more stringent than the $8\TeV$ limit.
Finally, we estimated the discovery reach with $300\,\text{fb}^{-1}$.

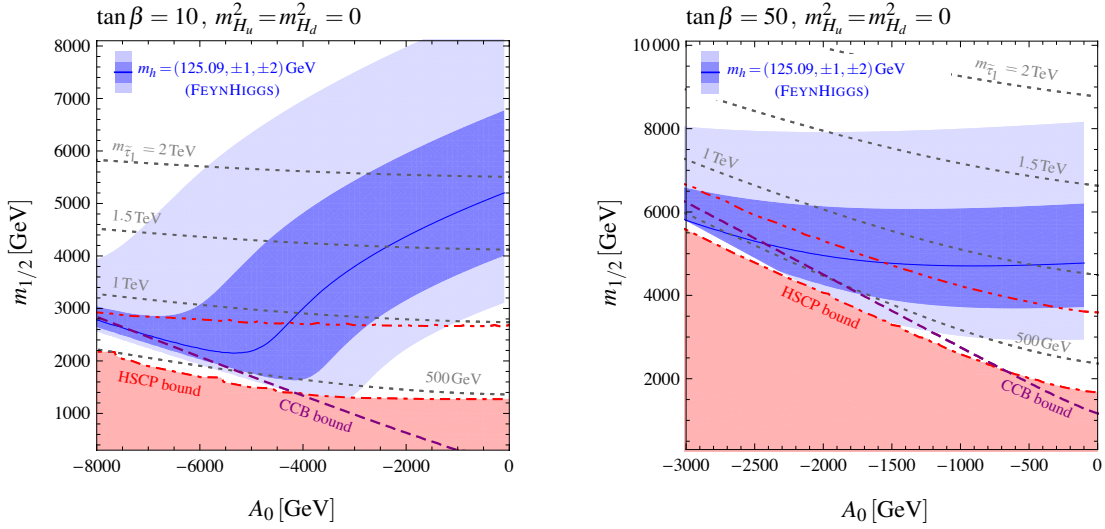
\begin{figure}[bt]
\centering
\setlength{\unitlength}{1\textwidth}
\begin{picture}(0.97,0.49)
 \put(-0.033,0.012){ 
  \put(0.07,0.03){\includegraphics[width=0.4\textwidth]{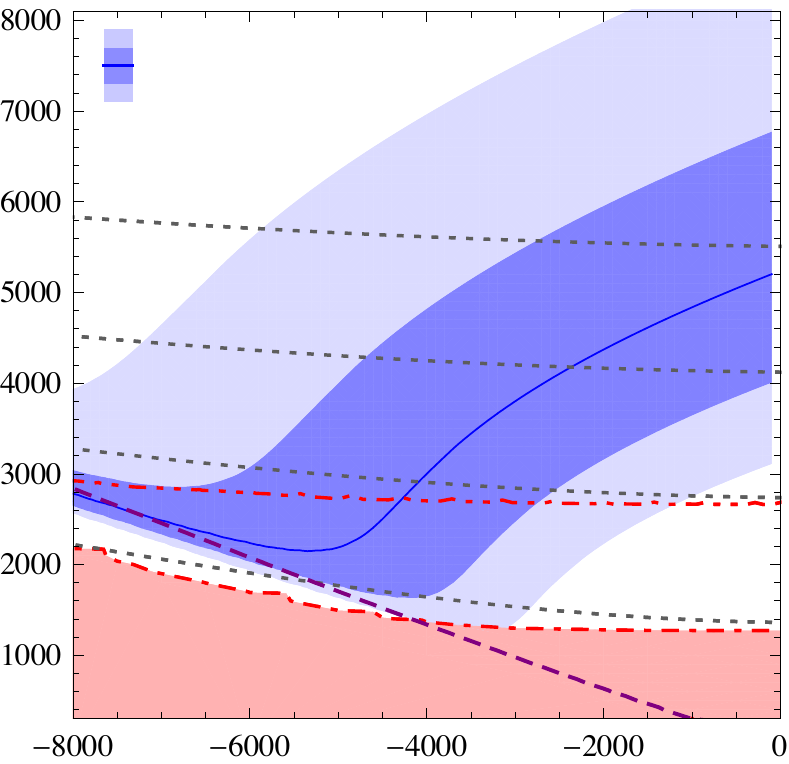}}
  \put(0.24,-0.005){\footnotesize $A_0 \,[\text{GeV}]$}
  \put(0.03,0.18){\rotatebox{90}{\footnotesize $m_{1/2} \,[\text{GeV}]$}}
  \put(0.143,0.3835){\rotatebox{0}{\tiny {\color{blue} $m_h\!=\!(125.09, \pm1,\pm2)\,\text{GeV}$}}}
  \put(0.184,0.364){\rotatebox{0}{\tiny {\color{blue} (\textsc{FeynHiggs})}}}
  \put(0.125,0.114){\rotatebox{-11}{\tiny  {\color{red} HSCP bound} }}
  \put(0.265,0.09){\rotatebox{-19.5}{\tiny {\color{ourpurple} CCB bound}} }
  \put(0.107,0.427){\rotatebox{0}{\footnotesize  $\tan\beta=10\,,\; m_{H_u}^2\!=\!m_{H_d}^2=0$}}
  \put(0.393,0.113){\rotatebox{-3.7}{\tiny {\color{gray} $500\,\text{GeV}$} }}
  \put(0.12,0.196){\rotatebox{-6}{\tiny {\color{gray} $1\,\text{TeV}$} }}
  \put(0.12,0.255){\rotatebox{-5.6}{\tiny {\color{gray} $1.5\,\text{TeV}$} }}
  \put(0.12,0.316){\rotatebox{-4.8}{\tiny {\color{gray} $m_{\widetilde\tau_1}=2\,\text{TeV}$} }}
  }
 \put(0.48,0.0117){ 
  \put(0.06,0.031){\includegraphics[width=0.409\textwidth]{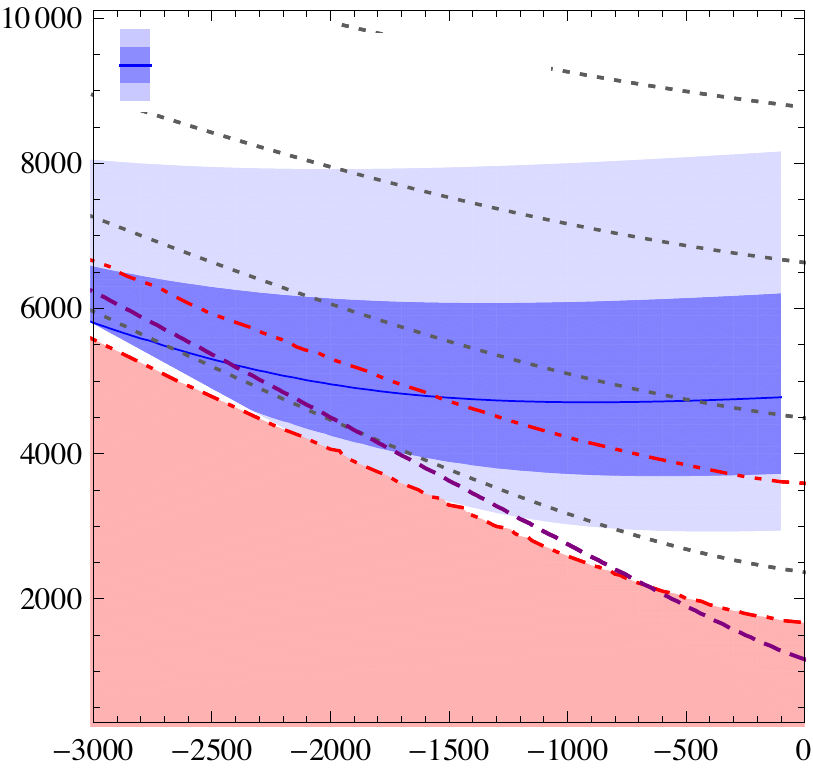}}
  \put(0.243,-0.005){\footnotesize $A_0 \,[\text{GeV}]$}
  \put(0.03,0.18){\rotatebox{90}{\footnotesize $m_{1/2} \,[\text{GeV}]$}}
  \put(0.143,0.3835){\rotatebox{0}{\tiny {\color{blue} $m_h\!=\!(125.09, \pm1,\pm2)\,\text{GeV}$}}}
  \put(0.184,0.364){\rotatebox{0}{\tiny {\color{blue} (\textsc{FeynHiggs})}}}
  \put(0.189,0.189){\rotatebox{-24.3}{\tiny  {\color{red} HSCP bound} }}
  \put(0.38,0.108){\rotatebox{-24.2}{\tiny {\color{ourpurple} CCB bound}} }
  \put(0.107,0.427){\rotatebox{0}{\footnotesize  $\tan\beta=50\,,\; m_{H_u}^2\!=\!m_{H_d}^2=0$}}
  \put(0.393,0.148){\rotatebox{-12}{\tiny {\color{gray} $500\,\text{GeV}$} }}
  \put(0.12,0.31){\rotatebox{-20}{\tiny {\color{gray} $1\,\text{TeV}$} }}
  \put(0.395,0.302){\rotatebox{-10}{\tiny {\color{gray} $1.5\,\text{TeV}$} }}
  \put(0.357,0.389){\rotatebox{-9.5}{\tiny {\color{gray} $m_{\widetilde\tau_1}=2\,\text{TeV}$} }}
  }
\end{picture}
\caption{
Constraints on the $A_0$--$m_{1/2}$ plane of gaugino mediation
with a stau NLSP and $\tan\beta=10$ (left) or $\tan\beta=50$ (right).
The grey dotted curves show the contours of a constant lighter stau mass
$m_{\widetilde\tau_1}$. 
The solid contours and blue-shaded bands show where the Higgs mass is
precisely $125.09\GeV$ and where it deviates from this value by at most
$1\GeV$ and $2\GeV$, respectively.
The red-shaded regions are excluded by searches for heavy stable charged
particles (HSCP) at the $8\TeV$ LHC\@.  For $\tan\beta=50$ and
$A_0\lesssim-2.3\TeV$ the HSCP limit extends into a region with a
tachyonic spectrum, where it is only an extrapolation.  The
dot-dot-dashed curves denote projections for $13\TeV$ and
$300\,\text{fb}^{-1}$.
The purple dashed lines indicate constraints from charge- and
color-breaking minima in the scalar potential.  Figures taken from
\cite{Heisig:2017lik}.
}
\label{fig:A0-M12-MH_constraints}
\end{figure}

The results are shown in \Figref{fig:A0-M12-MH_constraints} in the
$A_0$--$m_{1/2}$ plane for vanishing soft Higgs masses at the
compactification scale and two different values of $\tan\beta$.  In both
cases we have a stau NLSP in the entire considered parameter plane.  The
HSCP search by CMS \cite{Khachatryan:2015lla} at $\sqrt{s}=8\TeV$
excludes the red-shaded regions below the dot-dashed lines at 95\% CL\@.
The figure includes contours of constant stau mass to give an impression
of the physical mass spectrum.  Stau masses below about $400\GeV$ are
excluded.  For $\tan\beta=10$, this translates into a lower limit on
$m_{1/2}$ between $1$ and $2\TeV$; for large $\tan\beta$, this bound
becomes significantly stronger.  We do not show the preliminary run-$2$
limit to reduce clutter, but we include the estimated sensitivity for
$300\,\text{fb}^{-1}$ at $13\TeV$ as dot-dot-dashed curves.

Figure~\ref{fig:A0-M12-MH_constraints} also shows the contours on which
$m_h=125.09\GeV$ (blue solid lines) as well as the bands where the Higgs
mass deviates from this value by at most $1\GeV$ and $2\GeV$,
respectively (darker and lighter shaded regions).  We see that the
existing HSCP limit barely touches the $-2\GeV$ band, leaving most of
the parameter space with an acceptable Higgs mass unchallenged.  The
projected sensitivity curves indicate that future LHC runs will probe a
larger portion of this parameter space.  In particular, stau masses up
to almost $1\TeV$ can be tested.

In addition to collider searches, the emergence of charge- and
color-breaking (CCB) minima in the scalar potential for large trilinears
constrains the parameter space.
We applied the ``traditional'' condition for the stop trilinear coupling
\cite{Derendinger:1983bz,Casas:1995pd},
\begin{equation}
\label{eq:CCBtop}
A^2_t < 3 \left( \mhu^2+\mu^2+m^2_{Q_3}+m^2_{\bar{u}_3} \right) ,
\end{equation}
the analogous bound on the stau trilinear and the
upper limit on the product $\mu\tan\beta$ valid
for large $\tan\beta$ \cite{Kitahara:2013lfa}.
These bounds are superimposed in \Figref{fig:A0-M12-MH_constraints},
where we show the most constraining one in each case as a purple dashed
line.  For large negative $A_0$, they extend into the part of the
parameter space compatible with the observed Higgs mass and are slightly
stronger than the $8\TeV$ HSCP bounds.  Note, however, that these CCB
constraints are not entirely reliable
\cite{Camargo-Molina:2013sta,Chowdhury:2013dka} and can therefore only
serve as indicators of regions that \emph{might} be excluded.

In the parameter space regions with a neutralino or sneutrino NLSP, the
scenario is probed by standard missing energy searches at the LHC\@.
Again utilizing \textsc{MadGraph5\_aMC@NLO} and \textsc{Pythia}~6 for a
Monte Carlo simulation, we added
\textsc{CheckMate}~1~\cite{Drees:2013wra} to test the signal against all
$8\TeV$ ATLAS searches implemented in this tool.  These searches
considered final states with a significant amount of missing transverse
energy in addition to jets or leptons.  We found that even for the
lightest superparticle spectra with $m_h > 123\GeV$, the signal lies
below the exclusion limits by at least an order of magnitude. 
Consequently, the LHC will most likely not be able to discover SUSY if
gaugino mediation with a neutral NLSP is realized in nature.

\section{Conclusions}

We have verified that with a slight generalization of the original
scenario, gaugino-mediated SUSY breaking allows for large trilinear
scalar couplings, which help to lower the superparticle mass scale
required to reach the observed Higgs mass.  As the trilinears are
proportional to Yukawa coupling matrices, no new sources of flavor
violation arise.  Thus, the solution of the SUSY flavor problem in the
model is not endangered.

If the gravitino is the lightest superparticle and forms the dark
matter, the cosmological gravitino problem is alleviated.  In this case,
the next-to-lightest superparticle can be a neutralino or a slepton.
The observed Higgs mass is reached for $400\GeV \lesssim m_\text{NLSP}
\lesssim 1.4\TeV$.  As a consequence, a neutralino or sneutrino NLSP is
most likely too heavy for a discovery at the LHC, while a stau NLSP is
long-lived and potentially accessible in searches for heavy stable
charged particles.

\subsection*{Acknowledgements}
\noindent
J.H.\ acknowledges support by the German Research Foundation (DFG) through
the research unit ``New physics at the LHC''.
J.K.\ acknowledges financial support and hospitality from the Fine
Theoretical Physics Institute at the University of Minnesota and the
Meltzer Research Fund during the writing of these proceedings.

\newpage
\bibliographystyle{JHEP}
\bibliography{../../GauginoMediation}

\end{document}